\documentclass[%
twocolumn,
 longbibliography,
 amsmath,amssymb,
 aps,
 pra,
]{revtex4-1}

\usepackage{mathtools}
\usepackage{lipsum, color}
\usepackage{graphicx}
\usepackage{dcolumn}
\usepackage{bm}
\usepackage{comment}
\usepackage{afterpage}
\usepackage{hyperref}

\begin{document}
\title{ {Effective forces between active polymers}}
\author{M. C. Gandikota and A. Cacciuto}
\email{ac2822@columbia.edu}
\affiliation{Department of Chemistry, Columbia University\\ 3000 Broadway, New York, NY 10027\\ }

\begin{abstract}
\noindent
The characterization of the interactions between two fully flexible self-avoiding polymers is one of the classic and most important problems in polymer physics. In this paper we measure  these interactions in the presence of active fluctuations. We introduce activity into the problem using two of the most popular models in this field. One where activity is effectively embedded into the monomers’ dynamics, and the other where passive polymers fluctuate in an explicit bath of active particles. We establish the conditions under which the interaction between active polymers can be mapped into the classical passive problem. We observe that the active bath can drive the development of strong attractive interactions between the polymers and that, upon enforcing a significant degree of overlap, they come together to form a single double-stranded unit. A phase diagram tracing this change in conformational behavior is also reported.
\end{abstract}
\date{\today}
\maketitle

One of the defining features of active systems is their ability to transform energy in their environment into translational or rotational motion~\cite{dey_chemically_2017}.
This enhanced kinetics results in very exotic physical behavior not achievable in systems in thermal equilibrium. A significant amount of work has been done
to understand the interplay between active and thermodynamic forces (see for instance~\cite{zottl_emergent_2016,bialke_active_2015,menzel_tuned_2015,marchetti_hydrodynamics_2013,romanczuk_active_2012,bechinger_active_2016,cates_motility-induced_2015,marchetti_minimal_2016,ramaswamy_mechanics_2010,klapp_collective_2016,MalloryReview,wang_small_2013}) and to discover ways to exploit active forces to perform specific tasks at the micro scale~\cite{diLeonardo_bacteria_light,di_leonardo_active_2016}. Although most of the literature has focused on the behavior of spherical colloidal active particles, more recently, the way active particles interact with flexible and deformable objects or the behavior of active filaments has been the subject of intense scrutiny~\cite{Loi2011Oct,Kaiser2014Jul,Harder2014Dec,Ghosh2014Sep,Shin2015Oct,Kaiser2015Mar,Samanta2016Apr,Eisenstecken2016Aug,Chelakkot2014Mar,Isele-Holder2015Sep,Isele-Holder2016Oct,Kaiser2015Mar,Nogucci2016May,Bianco2018Nov,Harder2018Feb}
(see also \cite{Winkler2017Aug,winkler2020physics} for a brief review on the subject and references therein). Fully flexible active polymers are of particular interest. While there is model dependence for the scaling of the radius of gyration as a function of P\'eclet number~\cite{Bianco2018Nov},   the scaling behavior for polymers made out of active Brownian particles seems to remain unaffected by the action of active forces. Within this framework, the radius of gyration of a fully flexible active polymer follows Flory's exponent (at least for weak to moderate activities), and the active forces only affect the pre-factor of the scaling law~\cite{Kaiser2015Mar}. Similarly, we have recently shown~\cite{CacciutoDas2021} how the coil-to-globule transition of an active polymer with attractive interactions can also be understood with a rescaling of the temperature. Yet, some of its scaling behavior breaks down when polymers are placed under strong confinement~\cite{deviations2019}.

In this article, we study how adding activity changes the entropic forces two polymers exert on each other
within the framework of dry active matter. This is a classic problem in  polymer physics of passive polymers as it is of crucial importance to understand the phase behavior of dense polymer solutions~\cite{Gennes1979Nov,Khokhlov2002Mar}.
What is somewhat surprising about the interaction between two polymers in the infinite dilute limit is that the free-energy cost of fully overlapping two self-avoiding flexible polymers
is finite and accounts for only few $k_{\rm B}T$, where $k_{\rm B}$ is the Boltzmann's constant and $T$ is the temperature of the system~\cite{Grosberg82,Louis2000Sep}.

A simple way of rationalizing this result is to realize that overlapping two chains of $N$ monomers and radius of gyration $R^{(N)}_{\rm g}$, is similar to confining a
single chain of double the original length ($2N$) into a spherical cavity of radius equal to the radius of gyration of a single chain, i.e. $R=R^{(N)}_{\rm g}$. The free energy cost associated with this operation in units of $k_{\rm B}T$ is equal to $\beta F\sim (R_{\rm g}^{(2N)}/R)^{d/(d\nu-1)}$~\cite{Khokhlov2002Mar,Cacciuto2006May} where  $\beta=1/(k_{\rm B}T)$ is the inverse temperature, $\nu=3/(d+2)$ is the Flory scaling exponent and $d$ is the dimension of the embedding space. By plugging $R_{\rm g}^{(2N)}\sim (2N)^{\nu}$ and $R\simeq R^{(N)}_{\rm g}\sim N^{\nu}$, one obtains an estimate of the overlapping free energy $\beta F\sim 2^{d\nu/(d\nu-1)} $, which is clearly finite.
Unfortunately, this quantity cannot be directly measured in active systems as free energies cannot be consistently defined; yet, their
derivatives, i.e., pressures and forces the polymers exert on each other can be readily measured numerically.

There are two distinct models that have been put forward to study flexible active polymers in the context of dry active matter.
One where a passive chain is free to fluctuate in an explicit bath containing active particles~\cite{Harder2014Dec}, which we will refer to here as the ${\it explicit}$ model, and the other where the action of the active bath is incorporated into the chain by treating every single monomer as an effective independent Brownian active particle~\cite{kaiser_how_2015}, which we will refer to  as the ${\it implicit}$ model.

In experiments, the simplest realization of an active particle is obtained by coating one hemisphere of a silica or polymer micro-particle with a thin layer of platinum. Since the  metal hemisphere can be rather heavy, most active colloids readily deposit at the  bottom of the solution, and perform what is effectively a two dimensional Brownian active motion with the axis of propulsion parallel to the surface that supports them. We therefore limit our study to two dimensions, and when considering the explicit model, we envision the passive polymer as a chain of colloidal particles having the same diameter of the active colloids.

In this work, we will perform our measurements using both models.
Our basic model for a flexible, self-avoiding polymer consists of $N$ monomers of diameter $\sigma$ linearly connected with stiff harmonic springs and subject to thermal forces. Every monomer  undergoes Brownian dynamics at a constant temperature $T$. For the model with implicit active forces, a self-propelling force is introduced via a directional velocity of constant magnitude $v_p$  directed along a predefined orientation unit vector $\pmb{\hat{n}}$ centered at the origin of each monomer. For the model with explicit solvent, the monomers are exclusively subject to thermal forces, however, $N_a$ spherical particles of diameter $\sigma$ are also added in a simulation box of size $L$, and these follow the same active dynamics discussed above for the monomers of the implicit model.
The resulting equations of motion for both species of particles are
\begin{equation}
\begin{split}
\frac{d\pmb{r_i}(t)}{dt}  &=  \frac{1}{\gamma} \pmb{f}(\{r_{ij}\}) +   v_p \,  \pmb{\hat{n}_i}(t)\,\delta_{t_i,1}  + \sqrt{2D}\,\pmb{\xi}(t),\\
\frac{d \pmb{{\hat{n}_i}}(t) }{dt}&=\sqrt{2D_r}\, \pmb{\xi}_r(t) \times \pmb{\hat{n}_i}(t),
\end{split}
\end{equation}

\noindent  
where $i$ is the particle index, $t_i$ is a binary index that can acquire two values, 0 for passive particles and  1 for the active ones. $\delta_{t_i,1}$ is a Kronecker delta function which deactivates the self-propulsion term for the  passive particles. The translational diffusion coefficient $D$ is related to the temperature and the translational friction $\gamma$ via the Stokes-Einstein relation $D=k_{\rm B}T\gamma^{-1}$. Likewise, the rotational diffusion coefficient, $D_r=k_{\rm B}T\gamma_r^{-1}$, with $D_r = 3D\sigma^{-2}$. The  solvent induced Gaussian white-noise terms for both the translational $\pmb{\xi}$ and rotational $\pmb{\xi}_r$ motion are characterized by $\langle \pmb{\xi}(t)\rangle = 0$ and $\langle \xi_m(t) \xi_n(t^\prime)\rangle = \delta_{mn}\delta(t-t^\prime)$. $\pmb{f}(\{r_{ij}\})$ indicates the excluded volume forces for all particles and the harmonic forces between the monomers of each polymers.
Excluded volume forces between any two particles are enforced via a Weeks-Chandler-Andersen (WCA) potential
$U(r_{ij})=4\varepsilon\left[ \left( \frac{\sigma}{r_{ij}}\right)^{12} - \left(\frac{\sigma}{r_{ij}}\right)^{6} +\frac{1}{4}\right ].$
We use harmonic bonds between the monomers according to the potential  $U_b=k_b(r_{i,i+1}-\sigma)^2$. Here $r_{i,i+1}$ is the distance between consecutive monomers along the chain.
$k_b$ is set to $3500 k_{\rm{B}}T/\sigma^2$ to ensure polymer connectivity while simultaneously minimizing bond stretching that could arise from the action of the active forces. Finally, the hard repulsion between the monomers was selected to be $\varepsilon=100 k_{\rm{B}}T$, to prevent interpenetration of the polymers.  
To confine the distance between the center of mass of the two polymers, $\delta R_{\rm cm}$, to remain within a given distance $R$, we use a boundary defined by the  potential
\begin{equation}
U_c(\delta R_{\rm cm}-R) =\left \{\begin{array}{ll}
0,\quad &\delta R_{\rm cm}\leqslant R\\
k(\delta R_{\rm cm}-R)^2,\quad &\delta R_{\rm cm}> R\\
\end{array}
\right.
\label{constraint}
\end{equation}
with $k=2000 k_{\rm{B}}T/\sigma^2$.
Each monomer in a chain at position $\pmb{r}_i$ will experience a confining force
\begin{equation}\label{constraint2}
\pmb{f}_i=-\left(\frac{\partial U_c}{\partial\delta R_{\text{cm}}}\right)   
\frac{\partial \delta R_{\text{cm}}}{\partial \pmb{r}_i}     
\end{equation}
where $\pmb{\delta R}_{\text{cm}}=1/N\sum_{i=1}^N (\pmb{r}^{(1)}_{i}-\pmb{r}^{(2)}_{i})$, and the upper indices refer to whether the monomer is part of the first or the second chain.

In our simulations, $\sigma$ and $k_{\rm B}T$ are used as the units of length and energy scales of the system respectively, while $\tau=\sigma^2D^{-1}$ is the unit of time. All simulations were typically run for at least $10^9$ time steps with a time step ranging from $\Delta t=10^{-4}\tau$ to $\Delta t=10^{-5}\tau$. We skip the first $2\times10^6$ time steps to let the system achieve a state of dynamic equilibrium. The strength of the active forces is reported in terms of the P\'eclet number defined as $Pe=v_p\sigma/D$. The positions of the random active particles in the explicit model are randomly and uniformly distributed.

\begin{figure}[h]
\centering
\includegraphics[width=0.45\textwidth]{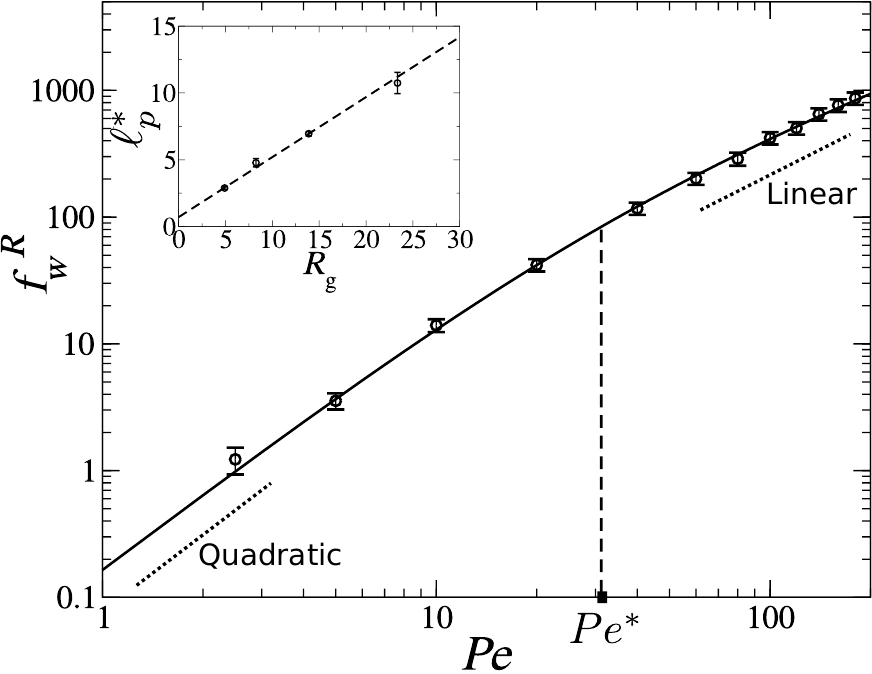}
 \caption{Reduced contact force $f_w^R$ as a function of $Pe$ for two polymers of length $N=256$.
 The vertical dashed line indicates the crossover point, $Pe^*$, between quadratic and linear behavior obtained as discussed in the text. The solid line is a fit to the data with the function $f(x)=ax/(1+b/x)$. The inset shows the linear relationship between $\ell_p^*=(Pe^*/3)\sigma$ and $R_{\rm g}$ for  $N=32,64,128, 256$.}
 \label{fig1}
\end{figure}

We begin our analysis with the ${\it implicit}$  model. We measure how the force exerted on the boundary, $f_w=-\partial U_c/\partial \delta R_{\text{cm}}$, in the fully overlapping regime, i.e. when the distance between the centers of mass of two polymers is confined within  $R=\sigma$ from each other, depends on the strengths of the active forces, $Pe$.
The results are shown in Fig.~\ref{fig1}, where we report the reduced force $f_w^R=[f_w(Pe)-f_w(0)]/f_w(0)$ as a function of $Pe$ $-$ written this way $f_w^R$
is effectively equivalent to the reduced  pressure between the polymers. The data displays a quadratic behavior for small activities and a linear behavior for large $Pe$. The data for all polymer lengths considered in this study, $N=32,64,128,256$, is accurately described when fitted to the functional form $f_w^R=a Pe/(1+b/Pe)$, previously proposed to describe the pressure of active particles within a cavity as a function of $Pe$~\cite{mallory_anomalous_2014}. The crossover P\'eclet number, $Pe^*$, is estimated using the fitted parameter $b$.
Crucially, we find that the crossover occurs when the persistence length of the active force, defined as   $\ell_p=v_p/D_r=(Pe/3)\sigma$, becomes of the order of the radius of gyration of the polymers, $R_{\rm g}$. This result is important because it indicates that as long as $\ell_p$ is much smaller than $R_g$,  activity acts as an effective temperature $T\propto Pe^2$, and the phenomenology of the parent passive system can be easily extended to incorporate the active forces. In the opposite limit, such a mapping becomes inappropriate. The inset in Fig.~\ref{fig1} shows that the crossover $\ell_p^*$ tracks with $R_{\rm g}$ as  $\ell_p^*\approx R_{\rm g}/2$.
An analogous result was obtained for the {\it implicit} model embedded in three dimensions (see Fig.~S1 in the supplementary material).

Next, we measure the full force curve between two polymers, each of $N=128$ monomers, as a function of $R$. The force, $f_w$, contains two contributions, one due to the repulsion between the polymers, $f_p$, and the other, the ideal term, $f_{id}$, due to the overall motion of each polymer's center of mass independently of the presence of the other polymer.
Since we are exclusively interested in extracting the effective polymer-polymer interactions, we subtract the second contribution from the first and report  $f_p=f_w(R)-f_{id}(R)$. $f_{id}$ is computed in the same manner as $f_w$, but now the excluded volume interaction between monomers from two different polymers is turned off.

\begin{figure}[h]
\centering
\includegraphics[width=0.5\textwidth]{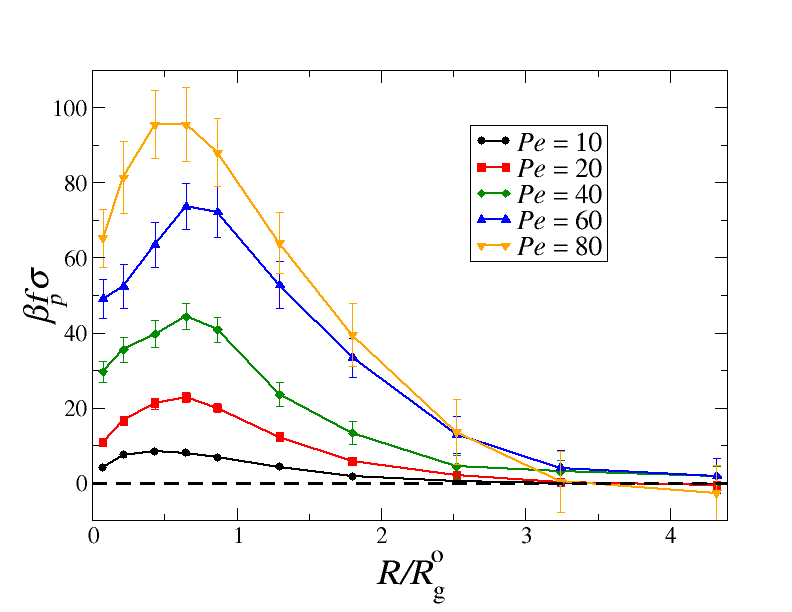}
 \caption{Force between two active polymers, $f_p$ as a function of their separation $R$, for different P\'eclet numbers. The data is normalized with respect to the radius of gyration of a passive polymer, $Pe=0$, of equal length, $R^{\rm o}_{\rm g}$.}
 \label{fig2}
\end{figure}

Figure~\ref{fig2} shows how $f_p$ depends on the polymer separation, $R$, for different values of $Pe$. The overall shape of the force as a function of $R$ is reminiscent of that expected for two passive polymers~\cite{Louis2000Sep}, where a steep repulsion observed as one moves from small to moderate overlaps leaves space to a decay of the force for large overlaps. This is consistent with a fully repulsive potential of mean force between the two polymers that flattens as the two polymers develop a significant degree of overlap.

\begin{figure}[h]
\centering
\includegraphics[width=0.5\textwidth]{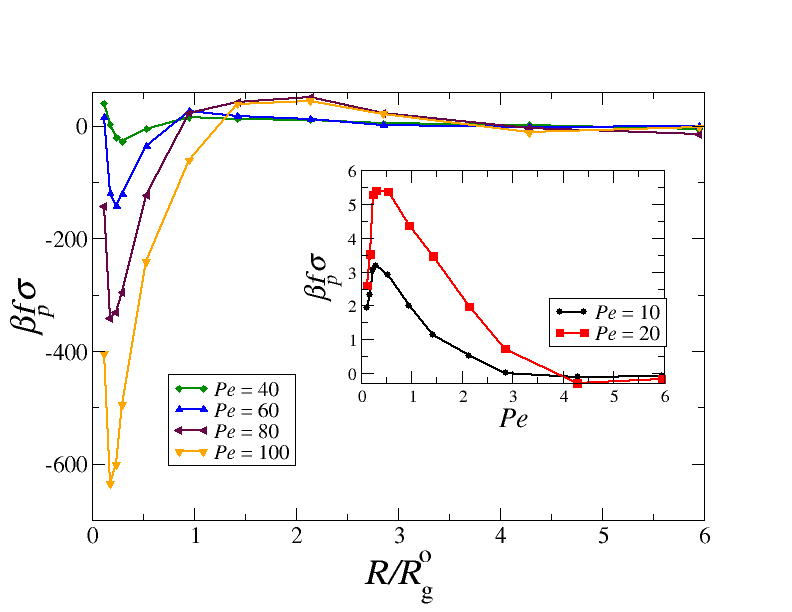}
 \caption{Force between two passive polymers, $f_p$, in a fluid of active particles, as a function of their separation $R$, for different P\'eclet numbers. The data is normalized with respect to the radius of gyration of a passive polymer, $Pe=0$, of equal length, $R^{\rm o}_{\rm g}$.
 The inset shows $f_p$ for small values of $Pe$.}
 \label{fig3}
\end{figure}

The phenomenology is richer for the {\it explicit} model. Here, we considered two $N=64$ passive polymers embedded in a square box of side length $L=100\sigma$ with periodic boundaries  containing $N_a=600$ active spherical particles. Each particle in the system interacts via the purely repulsive WCA potential discussed above.
This particular number density $\rho=N_a/L^2=0.06/\sigma^2$ is sufficiently large to ensure the induction of strong active fluctuations on the polymer, but not too large to drive motility-induced phase separation in the fluid. The net force between the polymers as a function of $R$, computed in a similar fashion as in the previous case \footnote{In this case $f_{\rm id}$ is computed by placing a single polymer inside a cavity of size $R$, confined via the truncated harmonic potential wall in Eq.~\ref{constraint} with $\delta R_{\rm cm}\rightarrow R_{\rm cm}$.} is shown in Fig.~\ref{fig3}. The data indicates a strikingly different behavior than what is observed for the {\it implicit} model.
For small P\'eclet numbers the functional form of the polymer-polymer force is similar for both models, however, beyond a certain P\'eclet number a net negative force develops for the {\it explicit} model, the depth and range of which increases with $Pe$. This is clear evidence of the development of an attractive interaction between the polymers at short range following a repulsion for intermediate separations.

This attraction develops as a result of condensation of active particles  along the perimeter of the polymers, which leads to an effective active depletion force not unlike that previously observed between rigid bodies in an active fluid  (see for instance~\cite{harder_role_2014,Ni2015Jan,Leite2016Dec,Baek2018Jan,Ray2014Jul}). The net effect is that at short separations the polymers pair up to form a single, double-stranded fluctuating unit. This is in opposition to the conformations in the {\it implicit} model, where such a pairing is not observed.

We should stress that the range, strength and sign of depletion interactions between passive objects immersed in active baths are known to be strongly shape dependent. In a two dimensional system of active depletants, passive colloidal plates experience an effective long-range attractive force whereas passive colloidal disks experience a short-ranged repulsive force~\cite{harder_role_2014}. Unlike rigid objects,
flexible polymers are free to fluctuate between shapes ranging from an extended rod-like conformation to a coiled up disk-like conformation.
Thus, the result that emerges from our simulations is rather non-trivial and neatly extends the results obtained for rigid objects to include fully flexible ones.

\begin{figure}[h]
\centering
\includegraphics[width=0.47\textwidth]{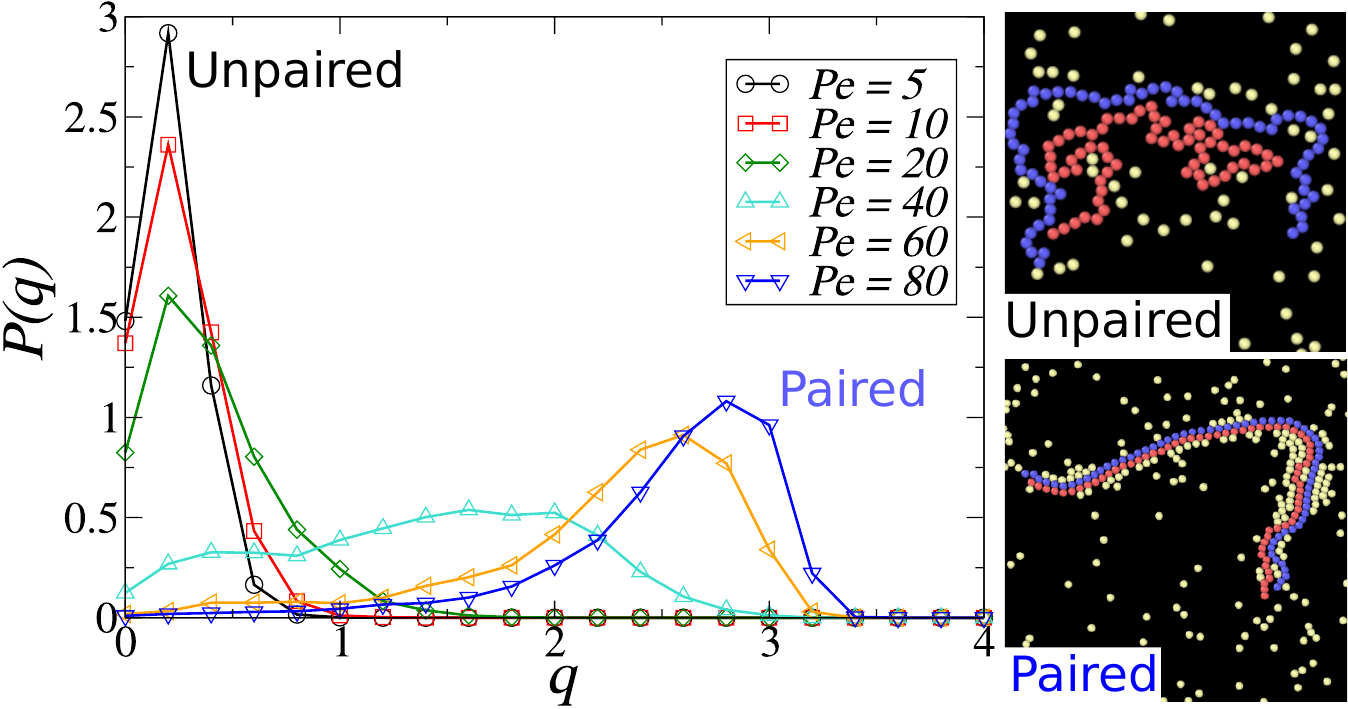}
 \caption{Left side: Probability distribution of the contact parameter $q$ as a function of P\'eclet number $Pe$. Right side: Snapshots from simulations showing characteristic unpaired (for small $Pe$) and paired (for large $Pe$) configurations of the two polymers in an active bath while their centers of mass are confined to be  within a distance of $R=\sigma$ from each other.   }
 \label{fig4}
\end{figure}

\begin{figure}[h]
\centering
\includegraphics[width=0.45\textwidth]{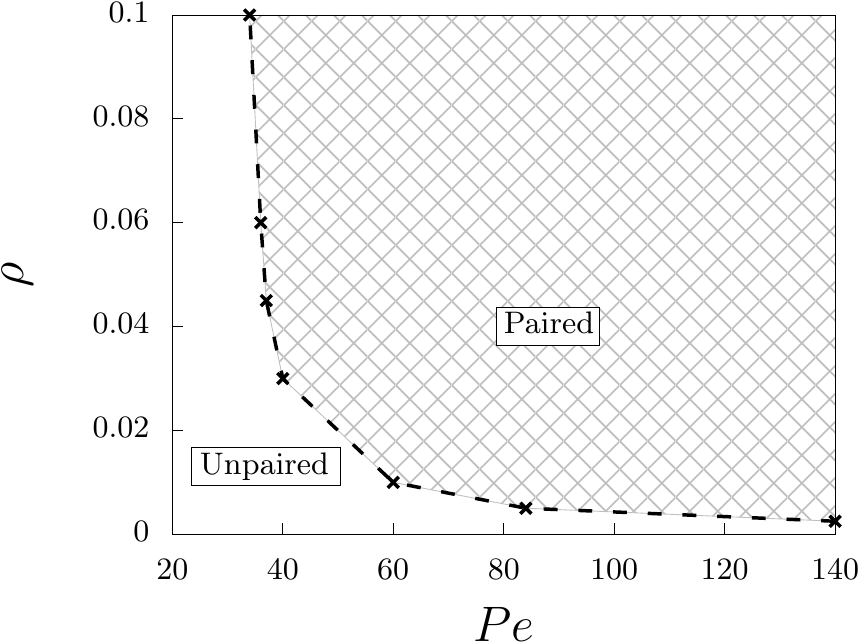}
 \caption{Phase diagram separating the paired and unpaired configuration of two polymers in an active bath with the distance between their centers of mass confined within a distance of $R=\sigma$ as a function of the density of active particles $\rho$ and P\'eclet number $Pe$. }
 \label{fig5}
\end{figure}

The right side of Fig.~\ref{fig4} shows typical paired and unpaired conformations associated with the two polymers at full overlap.
Although once in a while we observe some active particles in the interstitial space between the polymers, they can easily wiggle out because of their active forces and because of fluctuations between the polymer centers of masses that allow for some breathing room along the chain. It should be noticed that we also performed simulations with polymers of double the length ($N$=256) and even in this case we observe pairing of the polymer chains.
To characterize the transition between the two conformations as a function of the P\'eclet number, we measured the probability distribution of a contact parameter, $q$, defined as the average number of monomers in one polymer that are within a distance of $2\sigma$ from each monomer in the other polymer.
With this definition $q\in [0,4]$, where $q=0$ for fully unpaired polymers, and $q=4$ when they are fully aligned.
Figure~\ref{fig4} shows the shift of $P(q)$ as a function of  $Pe$ when the center of mass of the two polymers are confined to be within a distance of $R=\sigma$ from each other. A peak at small values of $q$ is visible for small $Pe$ and the peak moves closer to $q$'s maximum value for large $Pe$. In between, for $Pe\approx 40$, we see a broad distribution of $q$ indicating a  region where the polymer can move unobstructed from one conformation to the other.
Although the onset value of $Pe^{\dagger}\approx 40$ appears to be rather independent of $N$, apart from possible finite size effects, we do expect $Pe^{\dagger}$ to be very sensitive to the overall number density of active particles. By measuring  $P(q)$ for different values of $Pe$, we can construct a phase diagram tracing $Pe^{\dagger}$ for different values of active particles density $\rho$. The results are shown in Fig.~\ref{fig5}.
This diagram presents two interesting features. First, density and P\'eclet number are inversely related to each other, and it appears that even for very low densities it is possible to find a sufficiently large $Pe$ to drive the pairing of the polymers. Second, our data shows a lower bound for    
$Pe\approx 35$ below which no pairing occurs, even for large particle densities. We suspect this is because upon increasing the density, the number of active particles condensing on the two polymers become quite significant. The polymers act as crystallization seeds for the active particles that form multiple crystalline layers around and  between the contour length of the two polymers, effectively hindering their mobility. We find that the shape of the boundary is well described by the simple empirical equation $\rho=0.27/(Pe-31.3)$, which predicts a lower bound for $Pe\approx 31$.

\begin{figure}[h!]
            \centering
                \includegraphics[width=0.45\textwidth]{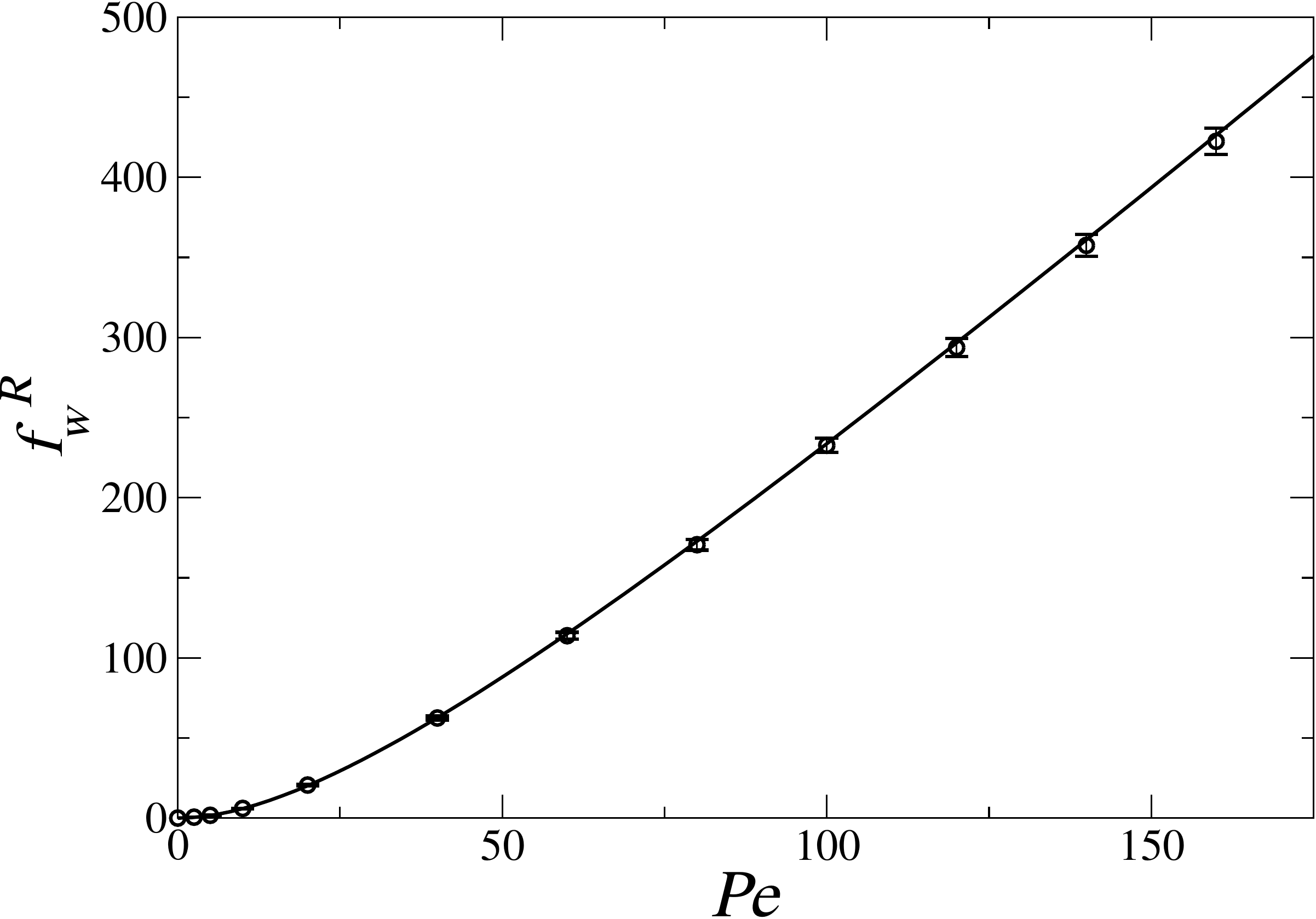}
                     \caption{Reduced contact force $f_w^R = [f_w(Pe)-f_w(0)]/f_w(0)$ as a function of Pe for two polymers of length $N=128$ in three dimensions. Here, both polymers are in the fully overlapping regime i.e. the distance between the centers of mass of the polymers are confined to be within a distance of $R=\sigma$, the monomer diameter. The reduced contact force moves from a quadratic regime at small Pe to a linear regime at large Pe. The solid line is a fit to the data with the function $f(x)=ax/(1+b/x)$. The fitting parameter $b=48.4$ is the $Pe^*$ at which the system moves across the two regimes. At this crossover, $l_p^*=Pe^*/3\approx2R_g$. Note that the pre-factor is different from the case of two polymers in two dimensions.}
                           \label{3D}
                       \end{figure}

Although we haven't performed systematic simulations of the full force-separation curves as a function of $Pe$ for polymers embedded in a three dimensional space, this is because most of the experiments with active colloids are in two dimensions, it is nevertheless of interest to look at the behavior in this case. Our sparse data in three dimensions for large P\'eclet numbers suggest that both the {\it implicit} (see Fig. \ref{3D}) and {\it explicit} models generate a purely repulsive interaction between the polymers which increases with the strength of the active forces. In this case the active depletion is not present as it is harder for the polymers to capture active particles along their contour length.

\indent In this article, we measured the forces between two active polymers. We used two different models to incorporate the role of active fluctuations into the  problem. In one model the fluctuations are implicitly incorporated into the motion of the monomers, in the other, they are explicitly accounted for by placing the polymers in an active bath of active particles. We find that in the former case, the forces between two fully overlapping polymers can be understood in terms of an effective temperature $T\propto Pe^2$ as long as $\ell_p$ is smaller than $R_{\rm g}/2$. Deviations from this behavior occur in the opposite limit where the forces grow linearly with $Pe$.

We observe a very different scenario when considering the explicit model. For small P\'eclet numbers, similar to the previous case, the net effect of the active bath is that of increasing the overall repulsion between the polymers. However, as soon as $Pe$ becomes sufficiently large for active particles to condense on the contour length of the polymers, a strong depletion attractive force emerges and drives the polymers to fluctuate as a pair.
Our results extend our current understanding of how active forces affect polymer fluctuations and set  clear limits for their mapping into effective equilibrium systems.
It is also important to emphasize that the emergence of active depletion force between fully flexible filaments is not a trivial result, and will have important implications for the dynamics and morphology of polymers solutions in an active bath.

An important limitation of our study is that it does not account for hydrodynamic effects~\cite{zottl_emergent_2016}. At a sufficiently large concentration of
active particles, hydrodynamic interactions could, in principle, drive orientational
instabilities that could destabilize the paired configurations.  
Whether this is the case requires more work in this direction, using a more sophisticated description of the active colloids with explicit squirmer models~\cite{Blake1971Mar} and explicit hydrodynamic interactions. We expect possible deviations to be dependent on the specific choice of swimming mode (puller vs pusher), and on whether the monomers in the chains are free to rotate or act as a fixed boundary against the torques applied by the squirmers~\cite{Das2019}.

\section*{Acknowledgements}
A.C. acknowledges financial support from the National Science Foundation under Grant No. DMR-2003444.

\bibliography{effective_forces} 
\bibliographystyle{rsc} 

\end{document}